\documentclass[aps,pra,reprint,floatfix,showpacs,twocolumn]{revtex4-2}
\usepackage{amsmath}
\usepackage[colorlinks=true,citecolor=blue,linkcolor=blue,urlcolor=blue]{hyperref}
\usepackage[usenames,dvipsnames]{xcolor}
\usepackage{longtable}
\usepackage{graphicx}
\usepackage{color}
\usepackage{float}
\usepackage{placeins}
\usepackage{rotating}
\usepackage{epsfig,rotate}
\usepackage{soul}
\usepackage{booktabs}
\usepackage{adjustbox}
\usepackage{xcolor}

\begin{document}

	\title{Dielectronic recombination studies on Fe$^{2+}$}

\author{S.~Singh}
\email{suvam.singh@mpi-hd.mpg.de}
\author{Z.~Harman}

\affiliation{Max--Planck--Institut f\"{u}r Kernphysik, Saupfercheckweg 1, 69117 Heidelberg, Germany}

	\begin{abstract}
	Dielectronic recombination resonance strengths, energy-differential cross sections, and recombination rate coefficients are calculated fully relativistically for Fe$^{2+}$ ions. The ground-state and resonance energies are determined using the multiconfiguration Dirac-Hartree-Fock method. Radiative and auto-ionization rates are computed with a relativistic configuration interaction method. For the calculation of Auger widths and resonance strengths, the continuum electron is treated within the framework of the relativistic distorted-wave model. Notably, the calculated level energies for Fe$^{2+}$ not only align well with experimental results but also show improvements compared to earlier theoretical studies. These fully relativistic calculations provide a more accurate and comprehensive understanding of the recombination process. This is particularly important in astrophysics and plasma physics, especially for studying phenomena such as kilonova events.

	\end{abstract}
	
	\date{\today}
	
	
\maketitle

\section{Introduction}\label{sect:intro}

The dielectronic recombination (DR) process \cite{Burgess1964DR} consists of two resonant steps. In the first step, an incident electron excites a bound electron during its recombination. The process is then completed by the radiative stabilization of the resulting autoionizing state. Such charge-state-changing processes play a crucial role in the dynamics of plasmas. These recombination processes also occur near the divertor walls of the magnetically confined fusion plasma in a tokamak reactor. The photons emitted during these processes provide valuable information about the plasma state. DR is the dominant mechanism for populating excited plasma states and inducing visible x-ray lines, which serve as diagnostic tools for fusion plasmas~\cite{Widmann95,beiersdorfer2015}. Additionally, DR significantly influences the energy balance of high-temperature fusion plasmas. Therefore, precise knowledge of DR processes is essential for accurately simulating fusion plasma reactors.

DR processes are also critical in astrophysics. They are vital for modeling electron absorption in interstellar clouds and for understanding the physics of outer planetary atmospheres. In both astrophysical~\cite{Massey42,Burgess1964DR,hitomi2016} and experimental plasmas~\cite{Cohen90,Cummings90}, DR serves as a powerful radiative cooling mechanism.

From a fundamental perspective, the state selectivity of DR experiments provides a stringent test of sophisticated atomic structure calculations. These tests are particularly important for investigating relativistic and quantum electrodynamics (QED) effects in bound electronic systems~\cite{Beiersdorfer2005,Brandau2003,Brandau2008}.

The recent detection of singly charged strontium \cite{watson2019identification} and doubly charged tellurium \cite{levan2023heavy} during a kilonova event highlights the need for more comprehensive investigations into the recombination of low-charged ions. This process significantly influences the ionization state and elemental composition of plasma \cite{muller2008electron}. With the recent launch of the James Webb Telescope (JWT), the potential to detect even more such events has increased. Dielectronic recombination involving low-charged atomic ions at low energies of the recombining electron remains insufficiently understood. Many of these systems are of significant astrophysical interest. X-ray satellite observatories, such as \textit{Chandra} and \textit{XMM-Newton}, have identified these elements as being of prime importance \cite{MclaughlinarXiv2012photoionization}. Iron (Fe) ions, for example, are present in the solar corona and the near-ultraviolet domain, where they account for approximately half of the total line-blanketing effect in cool stellar atmospheres. Their influence on the atmospheric structure of cool stars is significantly greater than that of the hydrogen Balmer lines \cite{Gehren1991Stellars}. Fe ions have also been detected in planetary nebulae \cite{bernitt2012unexpectedly,Kuhn2022}. Notably, Fe$^{2+}$ ions have been identified in the interstellar medium and in protoplanetary and planetary disks \cite{jager2016ion}. Ion-induced processing of cosmic silicates, such as amorphous MgFeSiO$_4$, plays a crucial role in the dust cycle within the interstellar medium and protoplanetary disks. This process can lead to the reduction of ferrous (Fe$^{2+}$) ions and the formation of iron inclusions in MgFeSiO$_4$ grains \cite{jager2016ion}.

\begin{figure}[h]
	\begin{center}		
		\includegraphics[width=0.9\textwidth, trim=0.5cm 5.5cm 0cm 0cm]{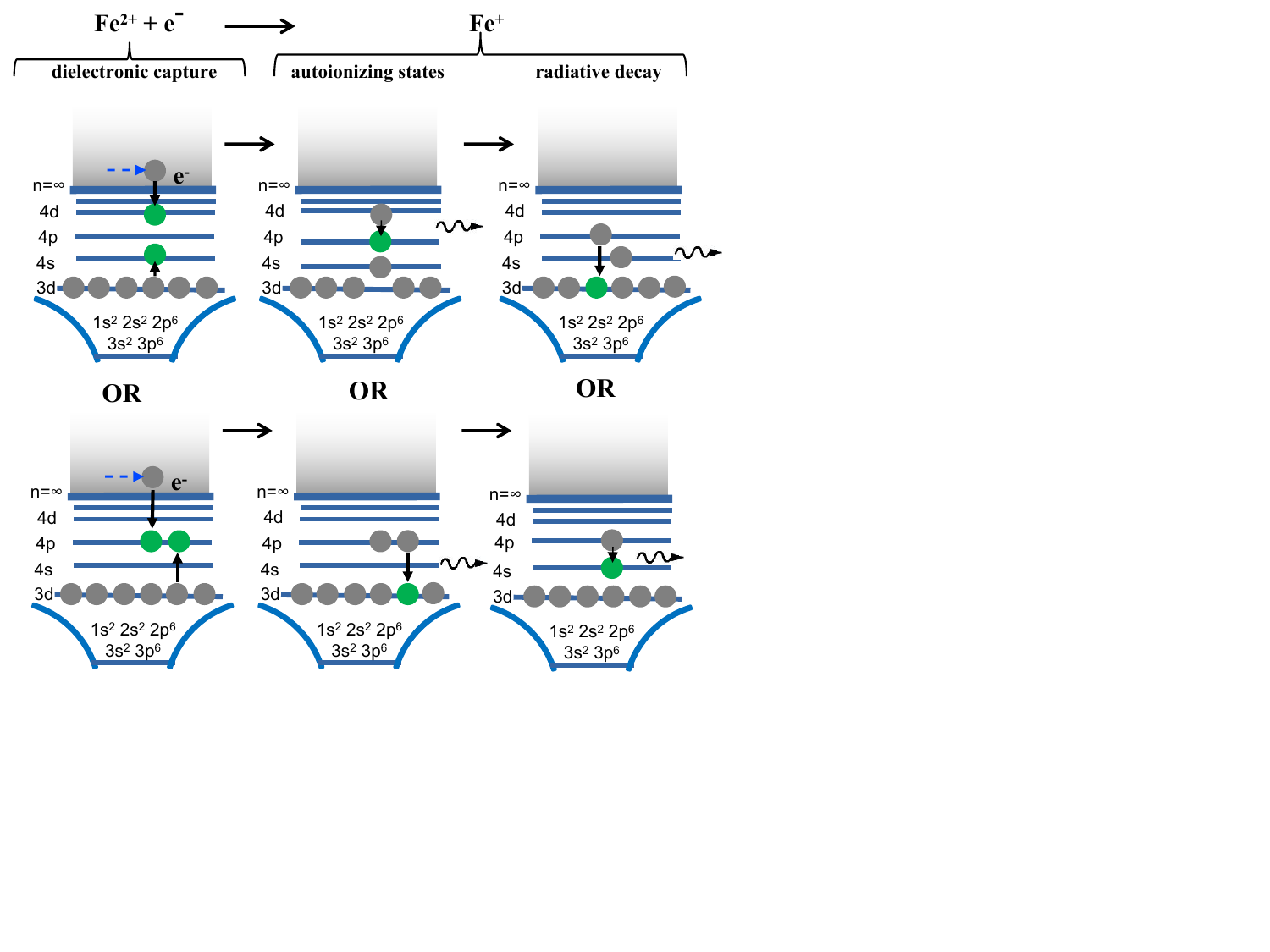}
		\caption{Schematic depiction of the DR process involving Fe$
		^{2+}$, showcasing the dielectronic capture, the autoionizing states, and the radiative decay process. The grey dots represent electrons in the initial state,
		while the green dots illustrate the movement of electrons to subsequent states.}\label{DR-schematic}
	\end{center}
\end{figure}

Studying DR in low-charged ions provides an opportunity to test accurate atomic structure models. Their structural complexity makes them an excellent test case, as they exhibit many possible excitation channels and multiple electron shells. By doing so, it is possible to establish how well the applied theoretical methods are reliable through a proper modeling of their DR processes. Numerous theoretical and experimental studies have explored parameters such as DR cross sections ($\sigma^{\mathrm{DR}}$), resonance strengths ($S^{\mathrm{DR}}$), and rate coefficients, particularly for highly charged system 
\cite{LindrothPRA2020, Lindroth-mahmood2012recombination, Lindroth2003high, fritzsche2009dominance, frtizscheandwu2015dielectronic, fritzsche2021dielectronic, badnell2018dielectronic, badnell2017recombination, gu2003dielectronic, GuWenBadnell2020rate,schuch2005dielectronic,xie2022calculations,luo2024linear}.
Although DR studies on Fe ions have received some attention, most have focused on highly charged Fe ions \cite{schippers2010dielectronic,gwinner2001dielectronic,griffin1987dielectronic,BrooksPRL1978}. A prior study by Nahar investigated Fe$^{2+}$ recombination \cite{nahar1997electron}, but their electron energy range of interest exceeded 10 eV. In contrast, our present investigation focuses on lower energies, specifically around 1 eV. 

The open $3d$ subshell renders the structure of Fe III ions complex.
It has many closely spaced energy levels, and electron correlation effects play an important role.
For Fe$^{2+}$, the states involved are:
\begin{eqnarray}
\label{eq1}
3d^6 ~(i) + e^- \to 3d^5 4s 4d ~(d)~ \& ~3d^5 4p^2~(d) \nonumber \\
\to 3d^5 4s 4p ~(f)~ \& ~3d^6 4p~(f)  \\
\to 3d^6 4s~(f') \nonumber
\end{eqnarray}
Here, \textit{i} denotes the initial state, which corresponds to the ground state of Fe III. The labels \textit{d} and \textit{f} represent the autoionizing state of Fe II and the final bound state of Fe II (with no further autoionization involved), respectively. Additionally, $f'$ denotes the ground state of Fe II. Figure~\ref{DR-schematic} provides a schematic representation of the various stages of the DR process considered in this work.

We simulate the low-lying resonance features by describing the ions using the multiconfiguration Dirac-Hartree-Fock (MCDHF) method \cite{Grasp2k,GRASP2018} and calculating cross sections with a distorted-wave technique. Bound states and transition energies are modeled using the GRASP2018 code package \cite{GRASP2018}. The auto-ionization and spontaneous decay rates, along with the atomic energy levels, are calculated using the Flexible Atomic Code (FAC) \cite{gu2003dielectronic,FACgu2004AIP}.

In the next section (Section~\ref{sect:theory}), the theoretical methodology is described in brief. Section~\ref{sect:results} describes the results and data analysis.
The paper concludes with a Summary (Section \ref{sect:summary}). Atomic units are used ($\hbar=m_e=e=1$), unless noted otherwise.

%
%
\section{Theory and calculation of resonance strengths}\label{sect:theory}
%
%
The complex structure of the involved states, especially that of the autoionizing and final levels necessitates an advanced many-electron model. The Dirac-Coulomb-Breit (DCB)
Hamiltonian of an atom or ion with $N$ electrons given by
\begin{equation}\label{eq:hdcb}
H^{\mathrm{DCB}}(\omega) = \sum_{i=1}^{N} h_i+\sum_{i<j}^{N} V_{ij}(\omega)\,,
\end{equation} 
with the single-particle operators
\begin{equation}\label{eq:hi}
h_i=c\vec{\alpha}_i \cdot \vec{p}_i +\left(\beta_i-1\right) c^2+ V_{\rm nuc}(r_i)\,,
\end{equation}
where $i$ and $j$ enumerate the electrons, and $c$ is the vacuum speed of light. The operator $\vec{p}_i$ is the relativistic momentum operator and $\vec{\alpha}_i$ and
$\beta_i$ are the Dirac matrices acting on the four-component wave function of a single electron.  $V_{\rm nuc}$ is the nuclear potential.

The electron-electron interaction operator $V_{ij}$ is given as the sum of the Coulomb and Breit interactions:
\begin{eqnarray}\label{eq:vij}
V_{ij}(\omega)   &=& V^C_{ij} + V^B_{ij}(\omega)\,,
\end{eqnarray}
with $V^C_{ij}=1/r_{ij}$ and $r_{ij}$ being the inter-electronic distance. The frequency-dependent Breit interaction $V^B_{ij}(\omega)$ is the sum of the following parts:
\begin{eqnarray}
V^B_{ij}(\omega)  &=& V^{\rm magn}_{ij} + V^{\rm ret}_{ij} + V^{\rm ret}_{ij}(\omega)\,.
\end{eqnarray}
Here, the Gaunt term is
\begin{eqnarray}\label{eq:magn}
V^{\rm magn}_{ij} &=& -\frac{\vec{\alpha}_i \cdot \vec{\alpha}_j}{r_{ij}}\,,
\end{eqnarray}
and the retardation term in the Breit approximation is~\cite{Breit29}
\begin{eqnarray}\label{eq:ret}
V^{\rm ret}_{ij}  &=& \frac{\vec{\alpha}_i \cdot \vec{\alpha}_j}{2r_{ij}} 
- \frac{(\vec{\alpha}_i \cdot \vec{r}_{ij})
(\vec{\alpha}_j \cdot \vec{r}_{ij})}{2r_{ij}^3}  \,.
\end{eqnarray}
The sum of $V^{\rm magn}_{ij}$ and $V^{\rm ret}_{ij}$ is the usual Breit interaction $V^B_{ij}$. Further retardation effects are accounted for by the term depending on the
frequency of the exchanged photon
$\omega$:
\begin{eqnarray}\label{eq:horet}
V^{\rm ret}_{ij}(\omega) &=& -\vec \alpha_i \cdot
\vec \alpha_j\frac{\cos(\omega r_{ij})-1}{r_{ij}} \\
&&+(\vec \alpha_i \cdot \vec \nabla_i)(\vec \alpha_j \cdot \vec \nabla_j)
\frac{\cos(\omega r_{ij})-1+\omega^2 r_{ij}^2/2}{\omega^2 r_{ij}}\,.\nonumber
\end{eqnarray}
In the above equation, $\vec \nabla_i$ stands for differentiation with respect to the coordinates of the $i$th particle.

In the MCDHF method~\cite{Gra70,Des71,Gra79,Gra88,Dya89,Par96,li2012mass}, an atomic state function is represented by a linear combination of configuration state
functions (CSFs) sharing the same total angular momentum ($J$), magnetic ($M$) and parity ($P$) quantum numbers~\cite{Gra79}:
\begin{equation}
\label{eq:asf}
|\Gamma P J M\rangle = \sum_{k=1}^{n_c} c_k |\gamma_k P J M\rangle\,.
\end{equation}
The $|\gamma_k P J M\rangle$ are built up as $jj$-coupled Slater determinants of relativistic orbitals. In Eq.~(\ref{eq:asf}), $\gamma_k$ summarizes all the information
needed to unambiguously describe the CSF, i.e. orbital occupation numbers and the angular momentum coupling scheme. The index $\Gamma$ denotes all the
$\gamma_k$ included in the representation of the atomic state function (ASF).

The quantum numbers of the single-electron orbitals are the principal quantum number $n$, the angular momentum $j=|\kappa|-\textstyle{\frac{1}{2}}$ and its magnetic
quantum number $\mu$:
\begin{equation}
\label{eq:orbitalwelle}
\phi_{n\kappa \mu}(\vec r)
=\frac{1}{r} \left( \begin{array}{c}
P_{n\kappa}(r) \Omega_{\kappa\mu}(\Omega) \\
i \, Q_{n\kappa}(r) \Omega_{-\kappa\mu}(\Omega)
\end{array} \right) \,.
\end{equation}
Here, $\kappa = 2(l-j)\,(j+1/2)$ is the Dirac angular momentum quantum number, while $P_{n\kappa}(r)$ and $Q_{n\kappa}(r)$ are the radial parts of the large and small
bispinor components. $\Omega_{\kappa m}(\Omega)$ is the spinor spherical harmonic.
The mixing coefficients in the ansatz (\ref{eq:asf}) are determined by diagonalizing the many-electron Hamiltonian with the given basis set, i.e., the $c_l$ are solutions
of the algebraic equation
\begin{equation}
\label{eq:coeffeq}
\sum_{l=1}^{n_c}
(\langle \gamma_k P J M | H^{\mathrm{DC}}| 
\gamma_l P J M \rangle-E_\Gamma^{\rm DC} \delta_{kl})
c_l=0 \,.
\end{equation}
As for the radial wave functions, the set of MCDHF equations given in ~\cite{Dya89} are solved iteratively together with the algebraic equations (\ref{eq:coeffeq}) until self-consistency is reached.

For a dielectronic recombination channel, i.e. for a two-step transition $i \to d \to f$, the cross section is expressed as a function of the electron kinetic energy
$E$ in the independent resonances approximation as (see, e.g.~\cite{HaanJacobs,Zim90,Zimmermann,Harman2019})
\begin{equation}
\label{eq:drkompakt}
\sigma^{{\mathrm{DR}}}_{i \to d \to f}(E) =
\frac{2\pi^2}{p^2}  V_a^{i\to d} \frac{A_r^{d \to f}}{\Gamma_d}  L_d(E).
\end{equation}
The initial state of the DR process consists of the ground-state ion and a continuum electron with an asymptotic momentum $\vec{p}$ and spin projection $m_s$.
In addition, $\Gamma_d$ is the total natural width of the intermediate autoionizing state, which is the sum of the radiative and autoionization widths:
$\Gamma_d = A^d_r + A^d_a$ (note that rates and corresponding line widths are equivalent in atomic units). Here, it is important to note that the subscripts $r$ and $a$ represent the radiative and autoionization processes, respectively. $L_d(E)$ is the Lorentzian line shape function, expressed as
\begin{equation}
\label{eq:lorentzian}
L_d(E) =  \frac{\Gamma_d/(2\pi)}
{(E_i+E-E_d)^2 +\frac{\Gamma_d^2}{4}},
\end{equation}
and is normalized to unity on the energy scale where $p=|\vec{p}|= \sqrt{(E/c)^2 - c^2}$ is the modulus of the free-electron momentum associated with the kinetic energy $E$.
The wave function of the continuum electron is represented by a relativistic partial wave expansion~\cite{Eichler},
\begin{eqnarray}
|E \vec{p} m_s\rangle&=&\sum_{\kappa m}i^l e^{i\Delta_{\kappa}} 
\sum_{m_l}Y_{l m_l}^*(\Omega) \\
&\times& C\left(l\ \frac{1}{2}\ j;m_l\ m_s \ m\right)| E \kappa m\rangle \,, \nonumber\label{Eq-Epms}
\end{eqnarray}
where $l$ and $m_l$ are the orbital angular momentum of the potential wave and the corresponding magnetic quantum number respectively. The phases $\Delta_{\kappa}$
are selected such that the continuum wave function satisfies the boundary conditions of an incoming plane wave and an outgoing spherical wave, as necessary for the
description of an incoming electron ({\em sic}, see Ref.~\cite{Eichler}). $\kappa = 2(l-j)\,(j+1/2)$ is the relativistic angular momentum quantum number. The total
angular momentum quantum number of the partial wave $| E \kappa m\rangle$ is $j=|\kappa|-\frac{1}{2}$. $Y_{lm_l}(\Omega)$ is a spherical harmonic and $\theta$ and
$\varphi$ denote the spherical angular coordinates. Here $C\left(l\ \frac{1}{2}\ j;m_l\ m_s\ m\right)$ represents the vector coupling coefficients.

The partial wave functions in the spherical bispinor form are expressed as
\begin{equation}
\psi_{E \kappa m}(\vec{r})= \langle \vec{r}| E \kappa m \rangle =
\frac{1}{r}
\left(\begin{array}{c} P_{E \kappa}(r)\Omega_{\kappa m}(\Omega)\\
i Q_{E \kappa}(r)\Omega_{-\kappa m}(\Omega)\end{array}
\right)\ .
\end{equation}
The index $d$ in Eq.~(\ref{eq:drkompakt}) denotes quantities related to the autoionizing state formed which constitutes the intermediate state in the
dielectronic capture process. This intermediate state then decays radiatively to the final state $f$. $V_a^{i \to d}$ denotes the dielectronic capture
(DC) rate and $A_r^{d}=\sum_f A_r^{d \to f}$ is the total radiative rate of the autoionizing intermediate state $|d\rangle$. The DC rate is given by
\begin{eqnarray}
\label{dr-rate}
V_a^{i \to d} &=& \frac{2\pi}{2(2J_i+1)} \sum_{M_{d}} \sum_{M_i m_s}
\int d\Omega \\
& & |\langle\Psi_{d}; J_{d} M_{d}  | V_C + V_B | \Psi_i E; J_i M_i, \vec{p}m_s\rangle|^2 \nonumber \\
&=& 2\pi \sum_{\kappa} |\langle\Psi_{d}; J_{d} || V_C + V_B 
|| \Psi_i E; J_i  j; J_d \rangle|^2 \,. \nonumber
\end{eqnarray}
In this equation, the matrix element of the Coulomb and Breit interaction\cite{Breit29} ($V^C$ and $V^B$, respectively) is calculated for the initial bound-free product
state $i$ and the resonant intermediate state $d$. After summation over the initial magnetic quantum numbers and integration over the direction $(\Omega)$ of the
incoming continuum electron, and after performing the summation over the magnetic quantum numbers of the autoionizing state, one obtains the partial wave expansion
of the reduced matrix elements, as given in the last line of the above equation.

The dielectronic capture rate is related to the rate of its time-reversed process, i.e., the Auger process, by the principle of detailed balance:
\begin{equation} 
\label{balance}
V_a^{i \to d} = \frac{2J_{d}+1}{2(2J_i+1)} A_a^{i \to d} \,.
\end{equation}
Here, $J_d$ and $J_i$ are the total angular momenta of the intermediate and the initial states of the recombination process, respectively. Neglecting the energy-dependence
of the electron momentum in the vicinity of the resonance, the dielectronic resonance strength, defined as the integrated cross section for a given resonance peak,
\begin{equation} 
S^{\mathrm{DR}}_{i \to d \to f} \equiv \int \sigma^{{\mathrm{DR}}}_{i \to d \to f}(E) dE\,,
\end{equation}
is given as
\begin{equation} 
S^{\mathrm{DR}}_{i \to d \to f}
= \frac{2\pi^2}{p^2} \frac{1}{2} \frac{2J_{d}+1}{2J_i+1} \frac{A_a^{i \to d}A_r^{d \to f}}{A_r^{d}+A_a^{d}}\,,
\label{eq:strength}
\end{equation}

where $A_a^{i \to d}$ is implicitly defined in Eq.~\eqref{balance}. The factor $\frac{2\pi^2}{p^2}$ defines the phase space density and the $1/2$ stems from the spin degeneracy
of the free electron. 

The total rate coefficients for thermal plasmas are described by the following equation, derived by summing across all possible autoionization channels and averaging over
the Maxwellian distribution of electron energies \cite{dubau1980dielectronic}. 
\begin{eqnarray} 
\label{rate-coeff}
\alpha_{\rm DR}(T) &=& \frac{h^3}{(2 \pi m_e k T)^{3/2}} \\
&& \sum_{d} \frac{2J_{d}+1}{2 (2J_{i}+1)} \frac{A_a^{i \to d}A_r^{d \to f}}{A_r^{d}+A_a^{d}} \exp \left(-\frac{E}{k T} \right)\,. \nonumber
\end{eqnarray}

In this expression, $k$ denotes the Boltzmann constant, $h$ is the Planck constant, $T$ represents the electron temperature, and $E$ is the resonance energy.

%
\section{Results}
\label{sect:results}

The MCDHF method, combined with the relativistic configuration interaction (RCI) approach, is used to calculate the energies of the auto-ionizing and radiative decay states of Fe$^{+}$, as well as the ground states of Fe$^{2+}$, as shown in Table \ref{transition_energies} and Table \ref{ground_state_energies}, respectively. The RCI calculations account for the Breit interaction and approximate QED corrections using the \textit{rci} code~\cite{GRASP2018}. To assess the impact of QED effects, we performed additional tests and found that, as expected for low-charged ions such as Fe$^{2+}$, these corrections contribute less than 0.5 meV. Given the present context, their influence on the final results is negligible.

\begin{table}
	\centering
	\begin{tabular}{lccc}
		\toprule
		Transition & Present & Nahar \cite{nahar1994atomic} & Johansson \cite{nave2012spectrum} \\
		&  &  &  and ref. therein \\
		\midrule
		$3d^64s$[$^6$D$_{9/2}$] $\rightarrow$  $3d^6$[$^5$D$_{4}$] & 15.9171 & 16.0302 & 16.1353 \\
		$3d^6$[$^5$D$_{4}$] $\rightarrow$ $3d^54s4d$[$^6$D$_{1/2}$] & 1.8298 & 1.9674 & 1.7633 \\
		$3d^6$[$^5$D$_{4}$] $\rightarrow$ $3d^54p^2$[$^6$S$_{5/2}$] & 1.7069 & & \\
		$3d^54s4p$[$^8$P$_{5/2}$] $\rightarrow$  $3d^6$[$^5$D$_{4}$] & 9.9064 & 10.5893 & 9.6562 \\
		$3d^64p$[$^4$F$_{9/2}$] $\rightarrow$  $3d^6$[$^5$D$_{4}$] & 10.5517 & 10.8410 & 10.6437 \\
		\bottomrule
	\end{tabular}
	\caption{Transition energies (in eV)}
	\label{transition_energies}
\end{table}

\begin{table}
	\centering
	\begin{tabular}{ccc}
		\toprule
		Config. state & Present (eV) & Ekberg (NIST) \cite{ekberg1993wavelengths,NIST_ASD} (eV)  \\
		\midrule
		$3d^6$[$^5$D$_4$] & 0 & 0  \\
		$3d^6$[$^5$D$_3$] & 0.0513 & 0.0541  \\
		$3d^6$[$^5$D$_2$] & 0.0871 & 0.0916 \\
		$3d^6$[$^5$D$_1$] & 0.1103 &  0.1156 \\
		$3d^6$[$^5$D$_0$] & 0.1218 &  0.1274 \\
		\bottomrule
	\end{tabular}
	\caption{Ground state energies (in eV) of Fe$
		^{2+}$}
	\label{ground_state_energies}
\end{table}

\begin{figure}
	\begin{center}		
		\includegraphics[width=0.69\textwidth, trim=2cm 6cm 0cm 0cm]{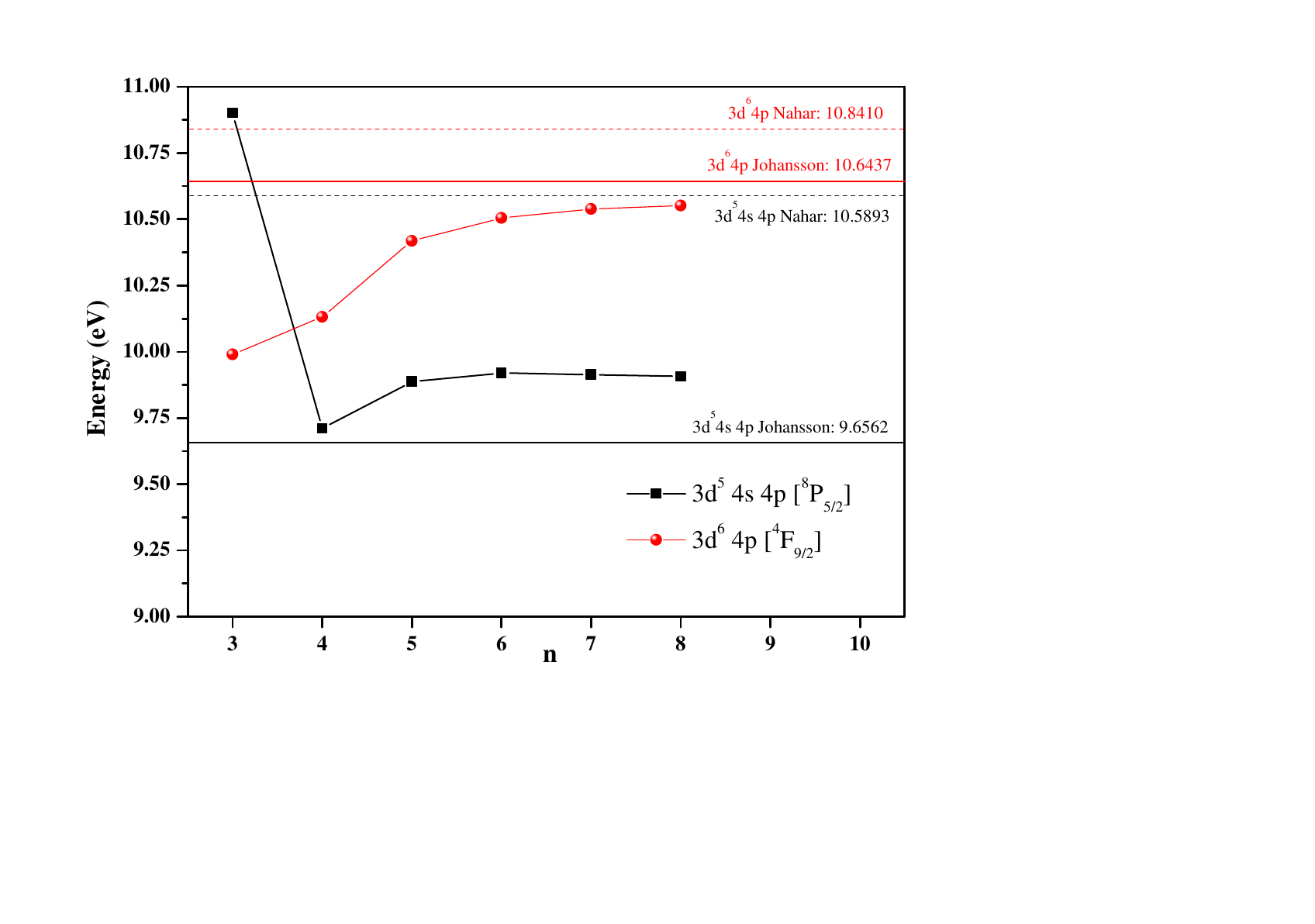}
		\caption{Calculated energies (in eV) for the states $3d^54s4p$[$^8$P$_{5/2}$] and $3d^64p$[$^4$F$_{9/2}$] monitored via layer-by-layer expansion. Black and red lines indicate present and previous studies done on the $3d^54s4p$[$^8$P$_{5/2}$] and $3d^64p$[$^4$F$_{9/2}$] states respectively.}
		\label{Levels-layer-by-layer}
	\end{center}
\end{figure}

Since electron correlations play a crucial role in this system, the ASFs are expanded using a large CSF basis. In the present calculation scheme, the CSFs are generated via single and double (SD) excitations of electrons from the Fe$^{2+}$ ($3d^6$ [$^5$D]) valence orbitals to auto-ionizing states of Fe$^{+}$ ($3d^54s4d$ [$^6$D] and $3d^54p^2$ [$^6$S]). A similar approach is applied for radiative decay states, where SD excitations of electrons from the $3d^6$ [$^5$D] valence orbitals to the radiative decay states $3d^54s4p$ [$^8$P] and $3d^64p$ [$^4$F] are considered. This method is also used to calculate transition energies from the ground state of Fe$^+$ ($3d^64s$ [$^6$D]) to Fe$^{2+}$ ($3d^6$ [$^5$D]). In these calculations, the spectroscopic orbitals in $3d$, $4s$, and $4p$ were kept open, depending on the states under consideration. This means that, during the generation of CSFs, electrons in these orbitals were allowed to undergo excitations to virtual orbitals as part of the CSF expansion. Valence-valence and core-valence correlations were included, and the convergence of the transition energies and radiative decay rates was monitored through a layer-by-layer expansion. The active space of virtual orbitals was optimized up to the $8k$ subshell. An example of this convergence is shown in Figure \ref{Levels-layer-by-layer}, where the energies of the  $3d^54s4p$ [$^8$P$_{5/2}$] and $3d^64p$ [$^4$F$_{9/2}$] states are seen to stabilize around the virtual orbital principal quantum number $n = 8$. We compare our result with the theoretical work of Nahar and Pradhan \cite{nahar1994atomic} and experimental work of S. Johansson (\cite{nave2012spectrum} and references therein), and it is observed that the present values agree well with the experimental value of S. Johansson (\cite{nave2012spectrum} and references therein) and demonstrate an improvement upon earlier theoretical work~\cite{nahar1994atomic}.

The number of CSFs for all the considered states is summarized in Table~\ref{tab:CSFs} along with the total angular momentum taken in the calculations as well.

\begin{table}
	\centering
	
		\begin{tabular}{l c c}
			\toprule
			State & J & Number of CSFs \\
			\midrule
			\( 3d^6 \) [\( ^5D \)]       & \(  0 \) to \( 4 \)    & 263,382       \\
			\( 3d^6 4s \) [\( ^6D \)]    & \(  1/2 \) to \( 9/2 \)    & 863,104       \\
			\( 3d^5 4s4d \) [\( ^6D \)]  & \(  1/2 \) to \( 9/2 \)    & 5,714,055     \\
			\( 3d^5 4p^2 \) [\( ^6S \)]  & \(  5/2 \)                & 957,456     \\
			\( 3d^5 4s4p \) [\( ^8P \)]  & \(  1/2 \) to \( 17/2 \)   & 5,986,145     \\
			\( 3d^6 4p \) [\( ^4F \)]    & \(  1/2 \) to \( 15/2 \)   & 2,955,693     \\
			\bottomrule
		\end{tabular}
	
	\caption{Number of CSFs considered for different states in the present calculations.}
	\label{tab:CSFs}
\end{table}

\begin{figure}
	\begin{center}		
		\includegraphics[width=0.66\textwidth, trim=2cm 3cm 0cm 1.5cm]{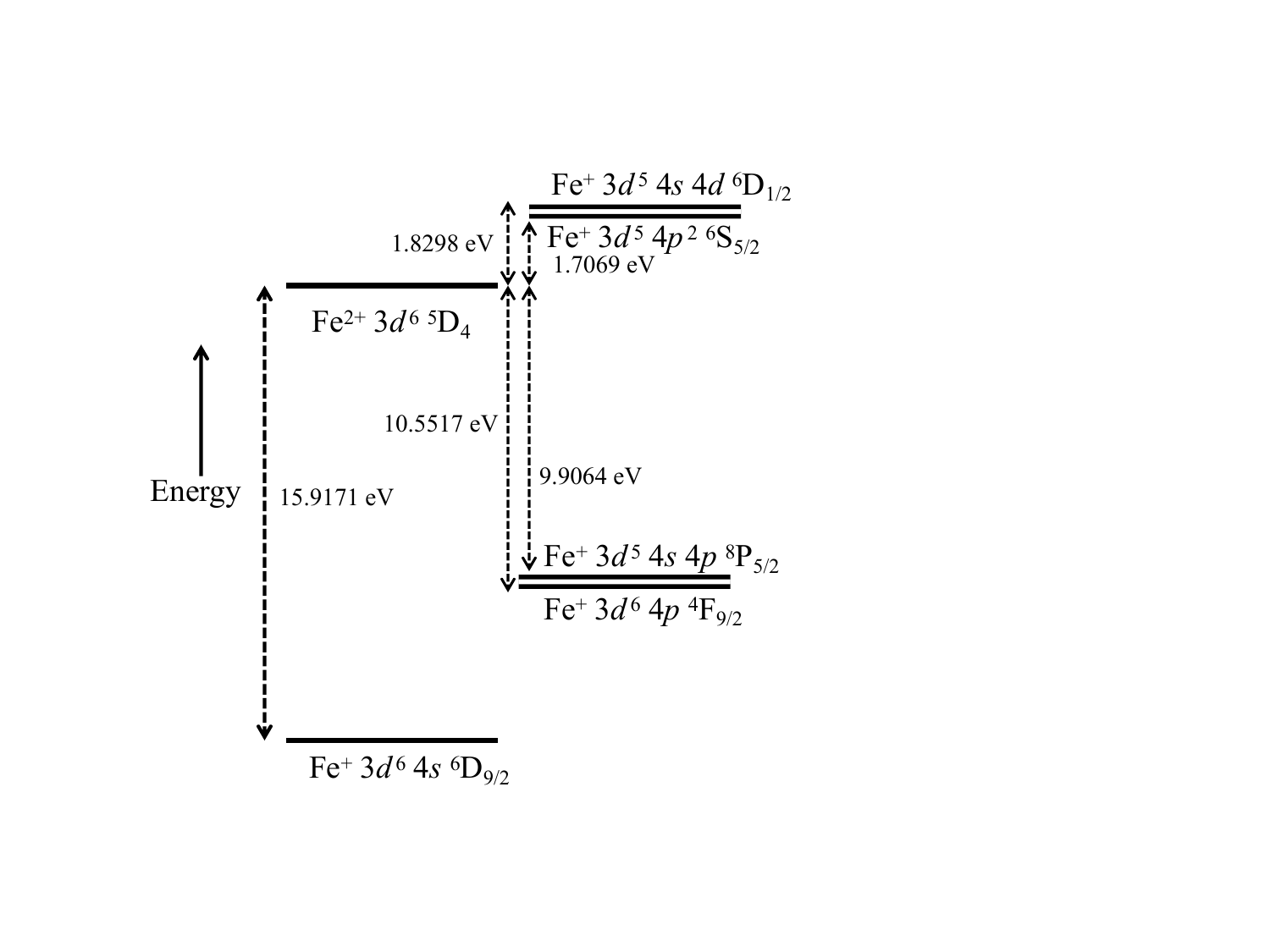}
		\caption{Calculated energy level diagram for the ground state of Fe$
			^{2+}$ and autoionizing and radiative levels of Fe$^+$. The energy axis is not to scale.}
		\label{Energy-level-Fe2}
	\end{center}
\end{figure}

The ionization energy of Fe$^+$ ($3d^64s$ [$^6$D$_{9/2}$]) to Fe$^{2+}$ ($3d^6$ [$^5$D$_{4}$]) is calculated to be 15.917eV, differing by only 1.3\% from the experimental values (see\cite{nave2012spectrum} and references therein). The autoionization level energies of the states  $3d^54s4d$ [$^6$D$_{1/2}$] and $3d^54p^2$ [$^6$S$_{5/2}$] from $3d^6$ [$^5$D$_{4}$] are calculated to be 1.830eV and 1.707eV, respectively, agreeing within 3.7\% of the experimental energies. Similarly, our calculated radiative decay levels of  $3d^54s4p$ [$^8$P$_{5/2}$] and $3d^64p$ [$^4$F$_{9/2}$] from $3d^6$ [$^5$D$_{4}$] are 9.9064eV and 10.5517eV, respectively, differing by 2.5\% and 0.8\% from experimental values. Since our calculations incorporate a more comprehensive set of relevant CSFs, the present results are expected to be more accurate than previous theoretical predictions \cite{nahar1994atomic}. These energy levels for various DR states are graphically represented in Fig.\ref{Energy-level-Fe2}. Table~\ref{transition_energies} provides a detailed comparison of these energy values with earlier studies. It is evident from the table that, apart from the  $3d^64s$ [$^6$D$_{9/2}$] $\rightarrow$ $3d^6$ [$^5$D$_{4}$] transitions, the present results show better agreement with experimental values compared to prior theoretical data. 

The ground-state energies for Fe$^{2+}$ are also calculated and presented in Table~\ref{ground_state_energies}. These values are compared with the experimental data provided by Ekberg \cite{ekberg1993wavelengths} in the NIST database \cite{NIST_ASD}. The calculated results exhibit excellent agreement with experimental data across all states. As discussed later, these energies were subsequently used in the FAC to generate accurate DR spectra. This underscores the enhanced accuracy and reliability of the current computational approach, demonstrating its capability to predict transitions and align well with experimental observations.

\begin{figure}[t]
	\begin{center}		
		\includegraphics[width=0.6\textwidth, trim=2cm 5cm 4cm 1cm]{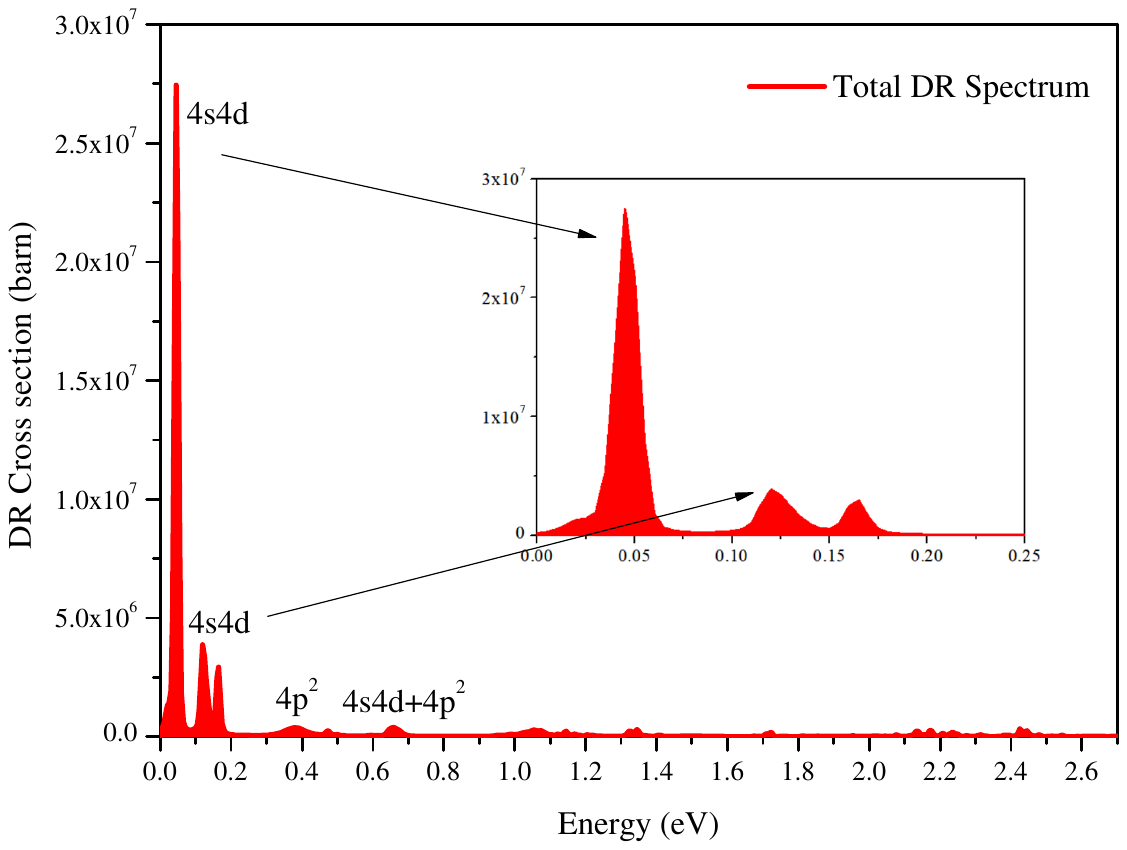}
		\caption{Total DR cross section for Fe$
			^{2+}$ recombining into Fe$^{+}$ is plotted against the relative electron energy.
			The Lorentzian line shapes (eq. ~\ref{eq:lorentzian}) are convoluted with a Gaussian function with a 10-meV width for a better representation of the spectra. The inset presents an enlarged view of the spectrum in the range from 0 to 0.25 eV.
		}
		\label{DR-spectrum-Fe2+}
	\end{center}
\end{figure}

We present the total DR cross section for Fe$^{2+}$ recombining to Fe$^+$, illustrated in Figure \ref{DR-spectrum-Fe2+}. The calculations employ the FAC code to derive the DR spectrum for two autoionizing states of Fe$^+$, namely, $3d^54s4d$ and $3d^54p^2$. To enhance the resolution, Lorentzian line shapes, as described in equation \ref{eq:lorentzian}, are convolved with a Gaussian function characterized by a width of 10~meV. The resulting spectrum exhibits pronounced resonances below 1 eV for both autoionizing states (especially for the $3d^54s4d$ state), with a gradual attenuation of the resonances. Notably, the contributions from the $3d^54s4d$ state are much stronger than from the $3d^54p^2$ state. We observe a significant resonance peak at around 0.05~eV, attributed to the $3d^54s4d$ state, along with a couple of resonance peaks between 0.1 to 0.2~eV. Beyond 0.2~eV, the spectrum is almost completely attenuated, with a small resonance peak at 0.4~eV due to the $3d^54p^2$ state. There is also a small resonance peak at 0.66~eV, resulting from the contribution of both autoionizing states $3d^54s4d$ and $3d^54p^2$. Since the energy levels calculated using the FAC code have an estimated uncertainty of a few eV, they are not highly accurate, necessitating adjustments to improve their precision. To achieve this, we replaced the FAC-calculated energies with more accurate values obtained from large-scale MCDHF calculations using the GRASP code. This substitution ensures more reliable energy positions for low-lying DR resonances. Further details on this energy adjustment procedure can be found in the FAC manual \cite{Gu_FAC}. While this approach enhances the accuracy of resonance positions, some uncertainty remains. In the present case, the ionization energy from Fe II to Fe III is calculated to be 15.9171 eV, which underestimates the experimental value by approximately 0.22 eV. This discrepancy suggests the possibility that the entire DR spectrum may be systematically shifted toward higher energies by a similar amount when compared to experimental data. Given that the deviations between theoretical and experimental values for various autoionizing states fall within this range, it is reasonable to infer that the overall uncertainty in the DR spectrum lies between the present theoretical prediction and a potential upward shift of approximately 0.22 eV. In addition to energy uncertainties, there are inherent uncertainties in the computed rates and cross sections. As noted by Gu in his FAC code \cite{Gu_FAC}, for near-neutral ions, such as Fe$^{2+}$, the uncertainties in rates and cross sections can be as large as 20\% or more. Therefore, it is reasonable to assume that the rates and cross sections in the present work have a minimum uncertainty of 20\%. The use of the MCDHF method in calculating resonance energies and radiative rates has proven effective in previous studies, yielding results that align well with experimental data \cite{Harman2019,baumann2014contributions,beilmann2011prominent}. This agreement underscores the validity of the theoretical approach employed in this study. Since a significant amount of data was generated in the present calculation, the energies, resonance strengths, along with their respective widths, are provided as supplementary data.

\begin{figure}[t]
	\begin{center}		
		\includegraphics[width=0.6\textwidth, trim=2cm 5cm 4cm 1cm]{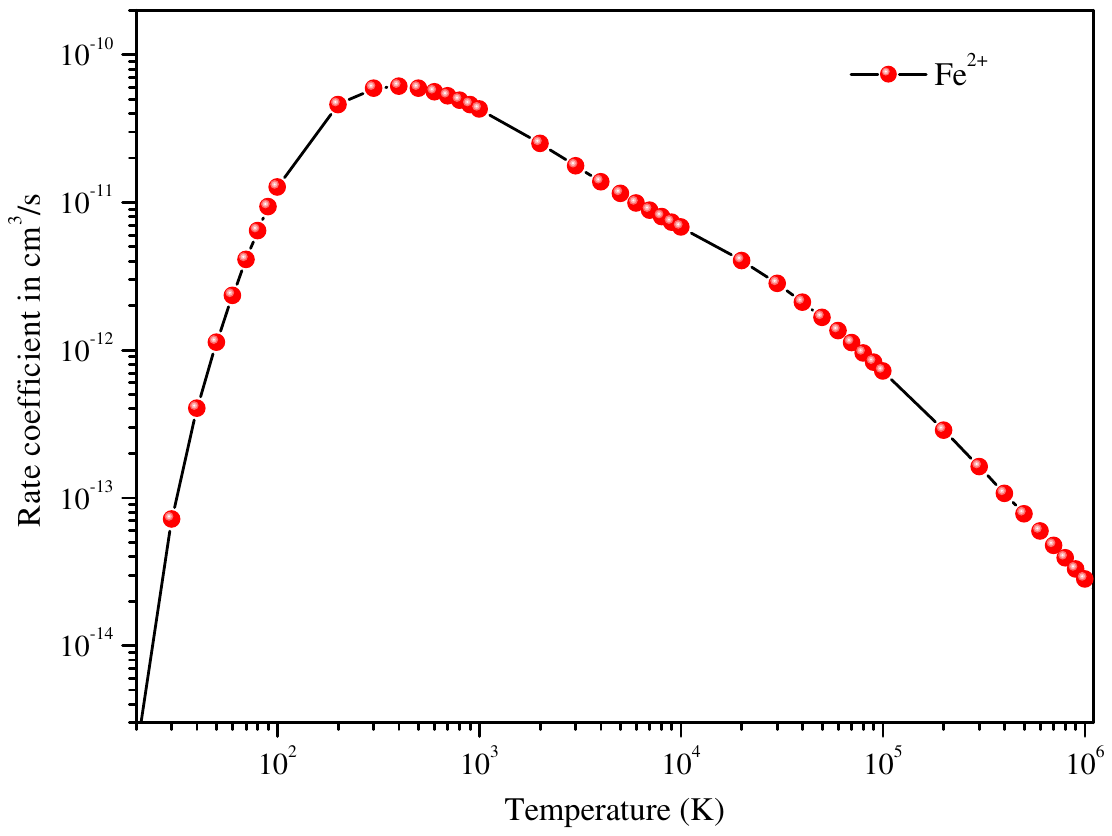}
		\caption{Total DR rate coefficients for Fe$
		^{2+}$ as a function of temperature.
		}
		\label{DR-rate-Fe2+}
	\end{center}
\end{figure}

The total DR rate coefficients for Fe$^{2+}$ are shown in the figure as a function of temperature on Fig. \ref{DR-rate-Fe2+}. At high temperatures, the DR rate coefficient
decreases according to a $\sim T^{-3/2}$ dependence. At low temperatures, however, the DR process is dominated by a few resonances located just above the ionization limit,
which contribute significantly to the rate at low temperatures. This occurs because the $\sim T^{-3/2}$ factor compensates for the low temperature. However, as the temperature
gets very low, the exponential cut-off eventually overwhelms the $\sim T^{-3/2}$ increase, causing the rate to decline sharply. As discussed above, the maximum uncertainty in the energy scale is approximately 0.22 eV, which corresponds to an uncertainty of about 2500 K in the temperature scale. Although this may appear significant at first glance, most of the DR rate coefficients contribute predominantly beyond the $10^4$ K temperature range \cite{kraemer2004effects}. Therefore, this level of uncertainty remains within acceptable limits. Given that the rate coefficients have a minimum uncertainty of 20\%, applying the error propagation rule from Eq. \ref{rate-coeff} yields a minimum overall uncertainty of approximately 28\% in the rate coefficient.

\section{Summary}
\label{sect:summary}

In this study, we conducted comprehensive calculations of DR spectra for recombination with Fe$^{2+}$ ions. This work may provide valuable guidance for future low-energy DR experiments on low-charged ions. The auto-ionization and radiative energy levels involved in the recombination process with Fe$^{2+}$, along with the ground-state energy of Fe$^{2+}$, were calculated. Our results align well with available experimental data and demonstrate an improvement over previous theoretical studies, underscoring the robustness and reliability of our methodology. These findings can also serve as a reference for future theoretical investigations.

Looking ahead, the theoretical framework established in this study is not limited to Fe$^{2+}$ ions. It can be extended to other systems, such as Sr$^+$, Y$^+$, Ce$^{2+}$, Te$^{2+}$, Xe$^{3+}$, and Xe$^{2+}$ ions, which are relevant to astrophysics, particularly in the context of kilonova events. The JWT offers an excellent opportunity to detect similar events in the near future. These many-electron ions exhibit multiple inner-shell excitation channels, resulting in the presence of different Rydberg series in the recombination spectra.

\bibliography{references}

\end{document}